# The Black Hole Information Puzzle and Evidence for a Cosmological Constant

Gerorge Chapline
Lawrence Livermore National Laboratory
Livermore, CA 94507

**Abstract**

Recent hints from observations of distant supernovae of a positive cosmological constant with magnitude comparable to the average density of matter seem to point in the direction of a two fluid model for space-time; where the "normal" component consists of ordinary matter, while the "superfluid" component is a zero entropy condensate. Such a two fluid model for space-time provides an immediate and simple explanation for why information seems to be lost when objects fall into a classical black hole.

hep-th/9807175  22 Jul 1998

Recent observations [1] of Type I supernovae at cosmologically significant redshifts have tended to confirm old suspicions [2] that the average density of matter is smaller than that required for a flat ($\Omega=1$) universe. In addition, these observations suggest that there is a positive cosmological constant whose magnitude is comparable to the average matter density. This last result is very surprising from the point of view of theoretical expectations based on either conventional quantum field theories or the types of superstring models of elementary particles that have been intensively investigated during the past few years. On the other hand, the author proposed some time ago [3] a model for the vacuum state of quantum gravity, which requires that four dimensional space-time be conformally flat on macroscopic scales; i.e. on length scales large compared to the Planck length the geometry is flat but possibly the cosmological constant $\Lambda \neq 0$. The fundamental degrees of freedom in this theory correspond to the "anyons" of a supersymmetric SU($\infty$) gauge theory defined on a 2-dimensional surface, but with time replaced with an extra compact Euclidean dimension (or equivalently the 2-cycles of an asymptotically locally Euclidean 4-

dimensional space). Very recently it has been suggested [4] that the $N \to \infty$ limit of Euclidean SU(N) gauge theories can be identified with certain kinds of supergravity theories with a positive cosmological constant. Thus there are now theoretical grounds to suspect that a theory of quantum gravity like that of ref.3 actually has a positive vacuum energy. In fact because the model of ref.3 also requires that on average space be flat, in the absence of matter this vacuum energy must have the critical value $\rho_c = 3H^2/8\pi G$, where H is the usual Hubble constant.

The model for quantum gravity proposed in ref.3 has the additional feature that the ground state is a superfluid-like state whose order parameter $\psi$ can, following the hint of ref.4 that the vacuum energy is positive, be tentatively identified with the cosmological constant $\Lambda$

$$\Lambda = 3H^2 |\psi|^2. \tag{1}$$

As is usual in a theory with a condensate ground state [5] we expect that the order parameter of the condensate will slowly decrease as the entropy of the universe is increased. On the other hand if the entropy is not too large the universe will remain flat on macroscopic scales so that

$$\Omega_m + \Omega_\Lambda = 1, \tag{2}$$

where $\Omega_\Lambda = |\psi|^2$ and $\Omega_m = \rho_m / \rho_c$ is the ratio of the matter energy density to the critical density. In this letter we would first of all like to point out that the recent measurements of the brightness and redshift of distant supernovae can be taken as evidence in support of a condensate model for the vacuum state of quantum gravity with order parameter satisfying equations (1) and (2). Secondly, the assumption that the energy density associated with finite entropy represents collective excitations of the condensate vacuum leads to a very simple resolution of perhaps the most perplexing enigma of contemporary theoretical physics; namely, although to an outside observer information appears to be lost when objects fall into a classical black hole, to a freely falling observer nothing extraordinary appears to happen upon crossing the event horizon of a black hole [6].

Actually our proposed resolution of this paradox is closely related to the fact that the introduction of a condensate vacuum for quantum gravity also yields an explanation for an old cosmological puzzle: why is the observed entropy of the universe so low ? In particular the observed entropy of the universe is vastly smaller than what one would at least naively expect in any local field theory of gravity with innumerable short distance degrees of freedom. Of course local quantum field theories of gravity have many other difficulties, and one might think that the entropy puzzle is simply a reflection of these other difficulties. However, the observed entropy is also much smaller than what is expected in superstring theories whose ground states have continuous moduli. The author has previously noted [7] that the entropy puzzle suggests that in reality the universe is in a nearly pure quantum state (which necessarily has nearly zero entropy).

The entropy puzzle also makes it difficult to attribute the apparent classical behavior of the universe on very large scales to "decoherence", since loss of quantum information on cosmological scales would according to the Landauer principle [8] lead to the generation of an enormous amount of heat. Instead one must seek an explanation for the fact that the universe behaves like a classical system on large scales in the fact that a condensate state can exhibit classical-like collective excitations. Indeed, it follows from the same type of argument that was originally introduced by Feynman [9] to show that superfluid helium has very few low lying states that a condensate state like that used in ref.3 to represent the ground state of quantum gravity would have very few low-lying excited states, and that these low-lying excited states exactly correspond to the long wavelength modes expected in a classical supergravity theory.

If we accept the idea that the familiar space-time background of classical physics corresponds to a condensate vacuum, then it follows that the average entropy density in the universe must be relatively low; for otherwise, the entropy would overwhelm the order of the condensate state and destroy the classical background space-time. That is, given that there is a classical space-time background, then the entropy density must be sufficiently small to allow this background to exist. Furthermore, for small values of the entropy we expect that the energy density associated with the entropy should be describable as a dilute gas of supergravity modes. Thus for small entropy we can use a two fluid model for the source terms in Einstein's equations, such that the total energy density and pressure will be given by the sum of the energy densities and pressures from the vacuum and a gas of supergravity modes. If we demand that space-time remain flat, then the sum of the energy densities must satisfy equation (2). We would now like to point out that, apart from providing a natural explanation for the magnitude of the observed cosmological constant, this two fluid model for space-time also provides a natural and immediate resolution for the black hole information puzzle.

A crucial clue is provided by the fact that in the case of superfluid helium there is a limiting velocity for superfluid flow in a channel [5]. At the limiting velocity it becomes energetically possible in the rest frame of the fluid to create an elementary excitation at the boundary of the channel. If the elementary excitation has energy $\varepsilon$ and momentum p, then the condition for creation of an elementary excitation is

$$\varepsilon(p) + p \bullet v < 0, \tag{3}$$

where v is the superfluid velocity. The critical velocity for flow without dissipation will be given by

$$v_c = \min \{\varepsilon(p) / p \}. \tag{4}$$

In the case of superfluid helium the minimum in equation (4) corresponds to the creation of quantized vortices. In the context of our two fluid model for space-time an analogous claim would be that in a space-time with boundaries there is a critical Lorentz boost parallel to a boundary, such that dissipation will occur at the boundary for greater boosts. The critical boost would be determined by equation (4).

For the usual elementary particles the minimum velocity in (4) is simply the velocity of light c. However, if the spectrum of collective excitations of the ground state as a function of wavenumber has a minimum for some non-zero wavenumber, then the limiting boost might correspond to a velocity less than c. In the case of superfluid-like ground states similar to that of ref. 3, it is expected that the spectrum of collective excitations will have a minimum when the wavenumber of the excitation (=p/h) corresponds to the average separation between the minimal 2-cycles in the condensate vacuum. Such a minimum presumably corresponds to a stable soliton-like solution of the appropriate supergravity equations, where the non-zero wavenumber to be used in evaluating (4) corresponds to the inverse of the scale size for the soliton. The product of mass and scale size for soliton solutions of supergravity equations would clearly be smallest for extremal black hole-like solutions, many examples of which are already known (e.g. ref.10). In Planck scale units the product of mass and scale size for these extremal black holes would be of order unity. The limiting velocity implied by equation (4) will therefore be close to the speed of light, but perhaps slightly less.

The picture we thus arrive at is that in 4-dimensional space-time with boundaries the condensate ground state of quantum gravity is stable under Lorentz boosts up to some boost close to the speed of light. When this critical boost is reached microscopic black hole-like states are spontaneously produced and the information contained in classical coherent states begins to be lost. This picture is relevant to the black hole information puzzle because a) the stationary frame of reference of a freely falling observer is related to that of an outside observer by arbitrarily large Lorentz boosts as the freely falling observer approaches the horizon, and b) near the horizon the geometry of space-time approaches that of anti-de Sitter space-time, which is bounded by a time-like cylinder [11]. Although this boundary represents spatial infinity, it is only a finite light travel time away and energy and information can flow from any point in the space to the boundary and visa versa. Therefore the condensate wavefunction almost certainly knows about this boundary, and we can be confident that our above arguments about a critical Lorentz boost apply to objects nearing the horizon (although the relevant Lorentz boosts are not exactly parallel to the time-like boundary, for objects near the horizon they are approximately parallel).

Evidently the information lost from classical coherent states as they approach the event horizon will reappear as microscopic excitations at the boundary of anti-De Sitter space. This production of heat may be viewed as a special case of Landauer's principle [8], which states that erasure of information contained in coherent states always results in the production of heat. It is of course tempting to speculate that this heat production can be identified with Hawking's radiation [12]. However, in contrast with Hawking's explanation for why a black hole emits heat radiation, our explanation for the emission of heat radiation leaves no doubts concerning the possibility that the unitarity principle of quantum mechanics may be violated, because, as is the case with applications of Landauer's principle under ordinary laboratory circumstances, our mechanism for the production of heat merely represents the flow of information from macroscopic coherent states to microscopic states.

Finally, it may be worthwhile noting a curious similarity between our resolution for the information loss puzzle and the duality between bulk infrared degrees of freedom in anti-de Sitter space and ultraviolet degrees of freedom on the boundary pointed out by Susskind and Witten [13]. Two differences between our point of view and their result should be kept in mind, though. First, while the microscopic states in our picture are generated near the boundary of anti-de Sitter space, they are still bulk excitations. Secondly, in our picture the loss of information comes about as a result of the subtle interaction between a ground state for quantum gravity with *positive* cosmological constant and the anti-de Sitter near horizon geometry of a black hole.

Acknowledgment. The author is grateful for conversations with Chris Hull and John Schwarz

.